\documentclass[aps,pra, twocolumn, amsmath, amssymb, showpacs, superscriptaddress]{revtex4}

\usepackage{graphicx}
\usepackage{bm}

\begin{document}

\title{Atomic-motion-induced spectroscopic effects nonlinear in atomic density in a gas}

\author{V. I. Yudin}
\email{viyudin@mail.ru}
\affiliation{Novosibirsk State University, ul. Pirogova 1, Novosibirsk, 630090, Russia}
\affiliation{Institute of Laser Physics SB RAS, Pr. Lavrentyev 15-B, Novosibirsk, 630090, Russia}
\affiliation{Novosibirsk State Technical University, Pr. Karl Marks 20, Novosibirsk, 630073, Russia}
\author{A. V. Taichenachev}
\affiliation{Novosibirsk State University, ul. Pirogova 1, Novosibirsk, 630090, Russia}
\affiliation{Institute of Laser Physics SB RAS, Pr. Lavrentyev 15-B, Novosibirsk, 630090, Russia}
\author{M. Yu. Basalaev}
\affiliation{Novosibirsk State University, ul. Pirogova 1, Novosibirsk, 630090, Russia}
\affiliation{Institute of Laser Physics SB RAS, Pr. Lavrentyev 15-B, Novosibirsk, 630090, Russia}
\affiliation{Novosibirsk State Technical University, Pr. Karl Marks 20, Novosibirsk, 630073, Russia}
\author{O. N. Prudnikov}
\affiliation{Novosibirsk State University, ul. Pirogova 1, Novosibirsk, 630090, Russia}
\affiliation{Institute of Laser Physics SB RAS, Pr. Lavrentyev 15-B, Novosibirsk, 630090, Russia}
\author{S. N. Bagayev}
\affiliation{Novosibirsk State University, ul. Pirogova 1, Novosibirsk, 630090, Russia}
\affiliation{Institute of Laser Physics SB RAS, Pr. Lavrentyev 15-B, Novosibirsk, 630090, Russia}


\begin{abstract}
The interatomic dipole-dipole interaction is commonly thought to be the main physical reason for spectroscopic effects nonlinear in atomic density. However, we have found that the free motion of atoms can lead to other effects nonlinear in atomic density $n$, using a previously unknown self-consistent solution of the Maxwell-Bloch equations in the mean-field approximation for a gas of two-level atoms with an optical transition at unperturbed frequency $\omega^{}_0$. These effects distort the Doppler lineshape (shift, asymmetry, broadening), but are not associated with an atom-atom interaction. In particular, in the case of $nk^{-3}_0<1$ (where $k^{}_0=\omega^{}_0/c$) and significant Doppler broadening (with respect to collisional broadening), atomic-motion-induced nonlinear effects significantly exceed the well-known influence of the dipole-dipole interatomic interaction (e.g., Lorentz-Lorenz shift) by more than one order of magnitude. Moreover, under some conditions a frequency interval appears in which a non-trivial self-consistent solution of the Maxwell-Bloch equations is absent due to atomic motion effects. Thus, the existing physical picture of spectroscopic effects nonlinear in atomic density in a gas medium should be substantially revised.
\end{abstract}


\maketitle
\section{Introduction}
Collective effects are one of the most interesting and important issues in fundamental physics. With regard to laser spectroscopy of resonant gas media and high-precision atomic clocks, collective effects, such as the interatomic dipole-dipole interaction, distort the resonance lineshape (shift, broadening, asymmetry) \cite{Friedberg_1973,Lewis_1980,Lorentz_2011,Kazantsev_1967,Vdovin_1967,Vdovin_1969,Srivastava_1976,Fleischhauer_1999,Pfau_2002,Pfau_2005,Pfau_2009,Fofanov_2011,Friedberg_2011,Weller_2011,Sokolov_2011,Adams_2012,%
Javanainen_2014,Jenkins_2014,Sokolov_2014,Bettles_2015,Scully_2015,Scully_2016,Adams_2016,Jenkins_2016,Jenkins_2016_PRA,Sokolov_2016,Pfau_2017,Adams_2018,Sokolov_2020,Chang_2004,%
Ostermann_2012,Ostermann_2013,Kramer_2016,Henriet_2019,Qu_2019,Liu_2020,Cidrim_2021,Bettles_2016,Glicenstein_2020,Masson_2020,Andreoli_2021,Fayard_2021,Sierra_2021}. The need to use a quantum mechanical many particle formalism (for example, a many-atomic density matrix), significantly complicates the theoretical description of such effects. In this case, most theoretical papers investigate the limit where the collisional broadening of the line exceeds the inhomogeneous Doppler broadening caused by the free motion of atoms as this somewhat simplifies the mathematical calculations. In real experiments, this condition is satisfied either for strongly heated atomic cells or for an ensemble of laser-cooled atoms. However, the opposite case of significant Doppler broadening is studied quite rarely as atomic density effects are assumed to be small.

\begin{figure}[t]
\centerline{\scalebox{0.45}{\includegraphics{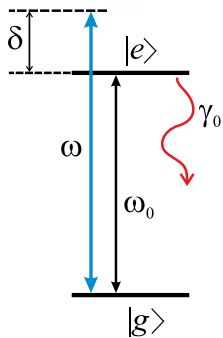}}}\caption{Two-level atom.} \label{two-level}
\end{figure}

As is well-known \cite{Lorentz_2011}, in an ensemble of two-level atoms with an unperturbed frequency $\omega^{}_0$ for a closed optical transition $|g\rangle\to |e\rangle$ (see Fig.~\ref{two-level}), the scale of the dipole-dipole interaction is determined by the value of Lorentz-Lorenz shift $\Delta^{}_{\rm LL}=-\pi nk^{-3}_0 \gamma^{}_0$, where $n$ is the atomic density (number of atoms per unit volume), $k^{}_0=\omega^{}_0/c$ is the wave number ($c$ is the speed of light in vacuum), $\gamma^{}_0$ is the spontaneous decay rate of the upper level (see Fig.~\ref{two-level}). In particular, for an ensemble of atoms confined within a layer of thickness ${\cal L}$, the total redshift induced by dipole-dipole interaction is (see Ref.~\cite{Friedberg_1973})
\begin{equation}\label{Sh_DD}
\Delta_{\rm dd}=\Delta^{}_{\rm LL}-\frac{3}{4}\Delta^{}_{\rm LL}\left(1-\frac{\sin 2k^{}_0 {\cal L}}{2k^{}_0 {\cal L}}\right)<0\,,
\end{equation}
where the second term is the collective Lamb shift. For a thick layer ($k^{}_0{\cal L}\gg 1$), the redshift (\ref{Sh_DD}) is equal to
\begin{equation}\label{DD}
\Delta_{\rm dd}= \frac{1}{4}\Delta^{}_{\rm LL}\approx -0.79nk^{-3}_0 \gamma^{}_0\,.
\end{equation}
However, when accounting for atomic motion in a gas, the results may differ substantially from Eq.~(\ref{Sh_DD}). For example, numerical calculations for large inhomogeneous broadening in Ref.~\cite{Javanainen_2014} show that the shift of Doppler lineshape significantly differs from the dependence (\ref{Sh_DD}) (see Fig.~2 (right panel) in Ref.~\cite{Javanainen_2014}). In the context of quantum microscopic theory, the interatomic dipole-dipole interaction is the result of the exchange of resonant photons (e.g., see Ref.~\cite{Sokolov_2011}). Therefore, atomic motion modifies the dipole-dipole interaction, since a photon emitted by one atom becomes non-resonant for an atom moving at a different velocity.

In addition to fundamental interest, effects nonlinear in atomic density are of great applied importance for high-precision laser spectroscopy and atomic clocks. In particular, the dependence of a reference resonance shift on atomic density $n$ determines the inaccuracy and long-term instability of atomic clocks due to the temperature variations, since the value of $n$ depends on the vapor cell temperature $T$.

In this paper, we describe spectroscopic effects nonlinear in the atomic density due to atomic motion, which follow from a self-consistent solution of the Maxwell-Bloch equations for atomic gases. These effects can significantly exceed the contribution of the interatomic dipole-dipole interaction to the distortion of the spectroscopic lineshape. In particular, the frequency shift of the Doppler lineshape for a monochromatic plane wave has an opposite sign and is more than one order of magnitude larger than the estimate (\ref{DD}). We emphasize that our results are obtained in the standard mean-field approximation without taking into account atom-atom interactions. The presented atomic-motion-induced effects have not previously been discussed in the scientific literature, as far as we know. Moreover, these effects cannot be described with a stochastic model of motionless atoms, in which inhomogeneous Doppler broadening is introduced by assigning to each atom a shift of the resonance frequency drawn at random from a Gaussian distribution.

\section{Basic formalism}

Let us consider propagation along the $z$ axis of a plane monochromatic wave with a real electric field $E(t,z)$ in a homogeneous gas medium of free-moving two-level atoms (see Fig.~\ref{two-level}). The atom-field interaction is described by the electro-dipole interaction operator $-\hat{d}E$. Our analysis is carried out within the framework of a self-consistent solution of the well-known Maxwell-Bloch equations, which include the wave equation for the field (in the CGS system):
\begin{equation}\label{Max_eq}
\left(\frac{\partial^2}{\partial z^2}-\frac{1}{c^2}\frac{\partial^2}{\partial t^2}\right) E(t,z)=\frac{4\pi}{c^2}\frac{\partial^2}{\partial t^2} P(t,z)\,,
\end{equation}
where $P(t,z)$ is the mean-field polarization of the medium. The atomic gas is described by the one-atomic density matrix $\hat{\rho}(v)$ ($v$ is the velocity of the atom) with the components $\rho^{}_{jl}(v)=\langle j|\hat{\rho}(v)|l\rangle$ (where $j,l=\{e,g\}$). In a linear approximation of the electric field $E$ (i.e., in the small saturation limit corresponding to the condition $|d^{}_{eg}E/\gamma^{}_{eg}|^2\ll 1$), we have the following Bloch equations:
\begin{eqnarray}\label{Bloch_eq}
&& \left(\frac{\partial}{\partial t}+v\frac{\partial}{\partial z}+\gamma^{}_{eg}+i \omega^{}_0\right) \rho^{}_{eg}(v) =id^{}_{eg} \rho^{}_{gg}(v)E(t,z)/\hbar\,, \\
&& \;\;\rho^{}_{ge}(v)=\rho^{\ast}_{eg}(v)\,,\quad \rho^{}_{gg}(v)=f(v)\,, \quad \rho^{}_{ee}(v)=0\,,\nonumber \\
&& \;\int^{\infty}_{-\infty}f(v)dv=1\,,\nonumber
\end{eqnarray}
where $\gamma^{}_{eg}$ is the optical coherence relaxation rate, $d^{}_{eg}=\langle e|\hat{d}|g\rangle=d^{\ast}_{ge}$ is the matrix element of the dipole moment operator, and the function $f(v)$ describes the velocity distribution of atoms. The operator $v({\partial}/{\partial z})$ in the left-hand side of Eq.~(\ref{Bloch_eq}) is a one-dimensional version (for the plane wave case) of the scalar operator $({\bf v}\cdot\nabla)$ for free-moving atoms in a gas, which is a fundamental difference from Bloch equations for an ensemble of motionless atoms where this operator is absent (e.g., for impurity resonant atoms in a solid). The polarization of a gas medium is defined as:
\begin{eqnarray}\label{P}
&& P(t,z)=n\langle D\rangle^{}_{v}\,,\\
&& \langle D\rangle^{}_{v}=\int^{\infty}_{-\infty}D(v)dv,\quad D(v)=d^{}_{ge}\rho^{}_{eg}(v)+c.c.\,,\nonumber
\end{eqnarray}
where $\langle D\rangle^{}_{v}$ is the velocity-averaged  dipole moment of the atom. Eqs.~(\ref{Max_eq})-(\ref{P}) constitute the Maxwell-Bloch system of equations in our case. Note also that we set the condition for the thickness of the gas medium ${\cal L}\gg\lambda$ to exclude the significant influence of various boundary effects \cite{Dicke_1953,Batygin_1953}.

In the case of a monochromatic plane wave traveling in a gas medium, the density matrix equations (\ref{Bloch_eq}) typically use $E(t,z)=[E_0\exp\{-i(\omega t-kz)\}+c.c.]$ (where $\omega\approx\omega^{}_0$ the light frequency, $k=\omega/c=2\pi/\lambda$ is the wave number in vacuum, $\lambda$ is the wavelength in vacuum), which leads to the standard expression for the Doppler lineshape. However, it is more correct to define the function $E (t,z)$ as a self-consistent solution of the equations system (\ref{Max_eq})-(\ref{P}), which we will seek in the form:
\begin{equation}\label{Etz}
E(t,z)=E_0e^{-i\omega t+Kz}+c.c.\,,
\end{equation}
where $K$ is an unknown complex number. Substituting the expression (\ref{Etz}) into right-hand side of Eq.~(\ref{Bloch_eq}), we find in the rotating wave approximation,
\begin{equation}\label{r_eg}
\rho^{}_{eg}(v)=\frac{id^{}_{eg}f(v)/\hbar}{\gamma^{}_{eg}-i\delta+Kv}\,E_0e^{-i\omega t+Kz},
\end{equation}
where $\delta=\omega-\omega^{}_0$ is the frequency detuning of the laser field (\ref{Etz}) in the laboratory reference frame. Then, using Eqs.~(\ref{r_eg}) and (\ref{P}), we obtain from Eq.~(\ref{Max_eq}) the following equation with respect to the unknown $K$:
\begin{equation}\label{K_eq}
K^2+k^2=-\frac{i4\pi k^2 n|d^{}_{eg}|^2}{\hbar}\left\langle\frac{f(v)}{\gamma^{}_{eg}-i\delta+Kv}\right\rangle_{v},
\end{equation}
where $\langle ...\rangle^{}_{v}$ denotes an integral over the velocity, $\int_{-\infty}^{+\infty}...dv$. Note that Eq.~(\ref{K_eq}) is obtained within the framework of a canonical approach. However, this equation and its solution have not been presented in scientific literature.

For the convenience in further analysis, we will make two transformations:

First, we represent the value of $K$ as follows
\begin{equation}\label{K}
K=(i+\alpha)k\,,
\end{equation}
where the dimensionless $\alpha$ describes the contribution of the gas medium. In this case, the correct (physical) solution must satisfy the condition ${\rm Re}[\alpha]<0$, which corresponds to the attenuation of the wave during propagation along the positive direction of the $z$ axis.

Second, using the well-known expression $\gamma^{}_0=4k_0^3|d^{}_{eg}|^2/(3\hbar)$ for the spontaneous decay rate of the upper level $|e\rangle$ (see Fig.~\ref{two-level}), we can represent Eq.~(\ref{K_eq}) as an equation with respect to the unknown $\alpha$:
\begin{equation}\label{alpha_1}
\alpha^2+2i\alpha=-i3\pi nk^{-3}_0 \gamma^{}_0\left\langle\frac{f(v)}{\gamma^{}_{eg}-i\delta+(i+\alpha)k\bar{v}v/\bar{v}}\right\rangle_{v},
\end{equation}
where we also introduced the parameter $\bar{v}$ characterizing the width of the velocity distribution $f(v)$. For example, in the case of the Maxwellian distribution, we have
\begin{equation}\label{fM}
f(v)=f^{}_M(v)=e^{-(v/\bar{v})^2}/(\bar{v}\sqrt{\pi})\;,\quad \bar{v}=\sqrt{2k^{}_BT/m}\;,
\end{equation}
where $k^{}_B$ is the Boltzmann constant, $T$ is the temperature of the gas, $m$ is the mass of the atom.

Since we are interested in the behavior of the lineshape under conditions close to the optical resonance ($\omega\approx\omega^{}_0$) in the range of several gigahertz, the magnitude of the relative change in the wave vector $\Delta k/k^{}_0$ is less than $10^{-4}$. Thus, we can safely use
\begin{equation}\label{kv}
k\bar{v}\approx k^{}_0\bar{v}=\Omega^{}_D\,,
\end{equation}
in the right-hand side of Eq.~(\ref{alpha_1}) to obtain the final equation
\begin{align}\label{alpha_2}
&\alpha^2+2i\alpha=\\
&-i3\pi nk^{-3}_0 \gamma^{}_0\left\langle\frac{f^{}_M(v)}{\gamma^{}_{eg}-i\delta+(i+\alpha)\Omega^{}_D v/\bar{v}}\right\rangle_{v},\; {\rm Re}[\alpha]<0,\nonumber
\end{align}
which we will use to determine the frequency dependence $\alpha(\delta)$ for each set of the fixed parameters $\{nk^{-3}_0,\gamma^{}_0,\gamma^{}_{eg},\Omega^{}_D\}$. It should be emphasized that although Eq.~(\ref{alpha_2}) was obtained in the framework of a one-atomic density matrix, there is the parameter $nk^{-3}_0 \gamma^{}_0$, which plays a key role in the theory of dipole-dipole interatomic interaction (i.e., in the case of many-atomic formalism).

\section{Atomic-motion-induced effects}
Note that if we remove the $\alpha^2$ contribution in the left-hand side of Eq.~(\ref{alpha_2}), the remaining equation
\begin{equation}\label{alpha_red}
\alpha=-(3/2)\pi nk^{-3}_0 \gamma^{}_0\left\langle\frac{f^{}_M(v)}{\gamma^{}_{eg}-i\delta+(i+\alpha)\Omega^{}_D v/\bar{v}}\right\rangle_{v},
\end{equation}
corresponds to the case when the reduced Maxwell equation is used instead of the full wave equation (\ref{Max_eq}). If we put $\alpha =0$ in the right-hand side denominator of the reduced equation (\ref{alpha_red}), we get the well-known expression for the Doppler-broadened lineshape,
\begin{equation}\label{Dopler}
\alpha(\delta)=-(3/2)\pi nk^{-3}_0 \gamma^{}_0\left\langle\frac{f^{}_M(v)}{\gamma^{}_{eg}-i\delta+i\Omega^{}_D v/\bar{v}}\right\rangle_{v},
\end{equation}
described by the standard Voigt profile, which is linearly proportional to the parameter $nk^{-3}_0\gamma^{}_0$. However, as follows from the basic equation (\ref{alpha_2}), $\alpha(\delta)$ should have a more complex nonlinear dependence on $nk^{-3}_0\gamma^{}_0$. Moreover, the presence of a nonlinear dependence on $nk^{-3}_0\gamma^{}_0$ follows even from the reduced equation (\ref{alpha_red}).

\begin{figure}[t]
\centerline{\scalebox{0.78}{\includegraphics{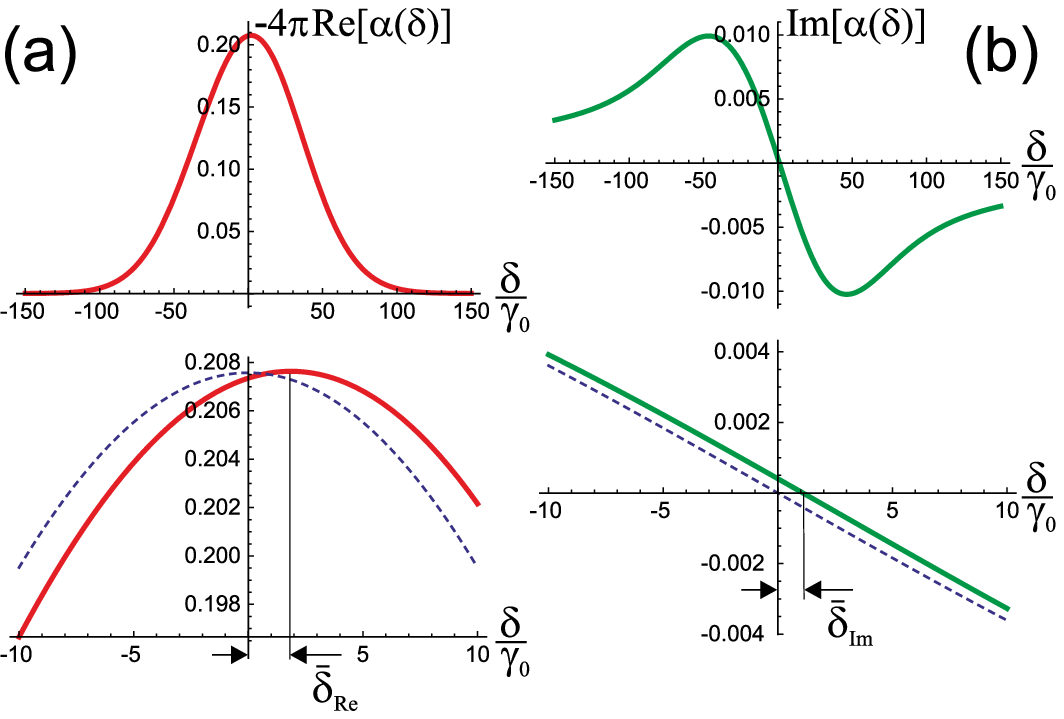}}}\caption{Spectroscopic dependence $\alpha(\delta)$ as a solution of Eq.~ (\ref{alpha_2}):\\
(a) the absorption coefficient Re[$\alpha(\delta)$] (red solid line); the bottom panel clearly shows the frequency shift $\bar{\delta}^{}_{\rm Re}$ in comparison to the standard Voigt profile (\ref{Dopler}) (see dashed line);\\
(b) the correction to the dispersion law Im[$\alpha(\delta)$] (green solid line); the bottom panel shows the frequency shift $\bar{\delta}^{}_{\rm Im}$ of the dispersion curve in comparison to the standard Voigt profile (\ref{Dopler}) (see dashed line). \\
The calculations are done with the following parameters: $nk^{-3}_0=0.1$; $\gamma^{}_{eg}/\gamma^{}_0=1/2$; $\Omega^{}_{D}/\gamma^{}_0=50$.} \label{Re_Im}
\end{figure}

Let us numerically investigate the solution of Eq.~(\ref{alpha_2}). Fig.~\ref{Re_Im}(a) shows the dependence of the absorption coefficient $4\pi{\rm Re}[\alpha(\delta)]$, where we use the factor $4\pi$ because the spatial dependence of the intensity $I(z,\delta)$ ($I\propto |E|^2$) during propagation in a medium has the form,
\begin{equation}\label{I}
I(z,\delta)=I^{}_0e^{4\pi{\rm Re}[\alpha(\delta)]z/\lambda}\,.
\end{equation}
As can be seen from Fig.~\ref{Re_Im}(a) (bottom panel), there is a positive shift $\bar{\delta}^{}_{\rm Re}$ for the top of the Doppler absorption lineshape. Our calculations show that under the conditions $\gamma^{}_{eg}/ \Omega^{}_D\ll 1$ and $nk^{-3}_0<0.5$ this shift is well described by the formula
\begin{equation}\label{delta_D}
\bar{\delta}^{}_{\rm Re}\approx 18.7\,nk^{-3}_0 \gamma^{}_0\,.
\end{equation}
In the Fig.~\ref{Re_Im}(b), we see the correction to the dispersion law ${\rm Im}[\alpha(\delta)]$, which is shifted by
\begin{equation}\label{delta_Im}
\bar{\delta}^{}_{\rm Im}\approx 11\,nk^{-3}_0 \gamma^{}_0\,.
\end{equation}
Comparing (\ref{delta_D}) and (\ref{delta_Im}) with the estimate (\ref{DD}), we can assert that in the case of $\gamma^{}_{eg}/\Omega^{}_D\ll 1$ and $nk^{-3}_0<0.5$ atomic-motion-induced effects are more than one order of magnitude greater than effects due to the dipole-dipole interaction. Moreover, shifts (\ref{delta_D})-(\ref{delta_Im}) have a positive sign (blueshifts), unlike the redshift (\ref{DD}).

Note that in Eq.~(\ref{alpha_2}), there are two sources of nonlinearity on the density parameter $nk^{-3}_0$ for $\alpha(\delta)$:\\
1) the quadratic contribution $\alpha^2$ in the left-hand side of Eq.~(\ref{alpha_2}), which is not related to atomic motion;\\
2) the term $\alpha\Omega^{}_D v/\bar{v}$ in the right-hand side denominator caused by the free motion of atoms in a gas.\\
In order to understand which of these factors is more significant, we use the reduced equation (\ref{alpha_red}), where the quadratic contribution $\alpha^2$ in the left-hand side is absent. In this case, we get the following results:
\begin{equation}\label{ReIm1}
\bar{\delta}^{}_{\rm Re}\approx 14\,nk^{-3}_0 \gamma^{}_0\,,\quad\bar{\delta}^{}_{\rm Im}\approx 7.5\,nk^{-3}_0 \gamma^{}_0\,.
\end{equation}
In addition, to clarify the role of atomic motion, consider the other equation
\begin{equation}\label{alpha_D}
\alpha^2+2i\alpha=-i3\pi nk^{-3}_0 \gamma^{}_0\left\langle\frac{f^{}_M(v)}{\gamma^{}_{eg}-i\delta+i\Omega^{}_D v/\bar{v}}\right\rangle_{v},
\end{equation}
obtained by ejecting $\alpha$ from the denominator in the right-hand side of Eq.~(\ref{alpha_2}). The physical solution of the quadratic equation (\ref{alpha_D}) has the form
\begin{equation}\label{alpha_D2}
\alpha(\delta)=-i\left[ 1-\left( 1+i\left\langle\frac{3\pi nk^{-3}_0 \gamma^{}_0f^{}_M(v)}{\gamma^{}_{eg}-i\delta+i\Omega^{}_D v/\bar{v}}\right\rangle_{v} \right)^{1/2}\right],
\end{equation}
which gives us the following result:
\begin{equation}\label{ReIm2}
\bar{\delta}^{}_{\rm Re}\approx 5\,nk^{-3}_0 \gamma^{}_0\,,\quad\bar{\delta}^{}_{\rm Im}\approx 5\,nk^{-3}_0 \gamma^{}_0\,.
\end{equation}
Comparing this with (\ref{delta_D})-(\ref{delta_Im}), we can assert that the atomic motion contribution [see the term $\alpha\Omega^{}_D v/\bar{v}$ in the right hand side denominator in Eq.~(\ref{alpha_2})] to nonlinear effects is dominate in the case of $nk^{-3}_0<0.5$ and $\gamma^{}_{eg}/\Omega^{}_D\ll 1$. However, it is interesting to note that even the shift (\ref{ReIm2}), caused only by the quadratic contribution $\alpha^2$ in the left-hand side of Eq.~(\ref{alpha_2}), is six times greater than the value (\ref{DD}).

\begin{figure}[t]
\centerline{\scalebox{1.2}{\includegraphics{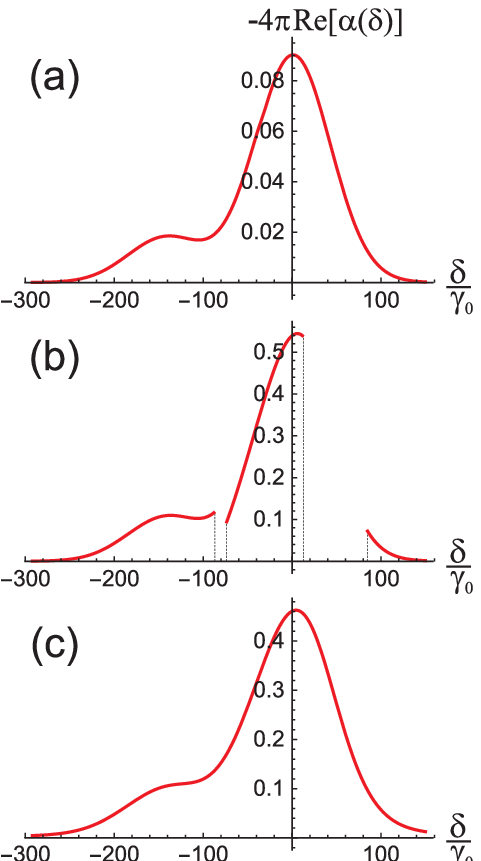}}}\caption{Frequency dependence of the absorption coefficient Re[$\alpha(\delta)$] for the $^{87}$Rb D1 line in the region of hyperfine transitions $F^{}_g=1\to F^{}_e=1$ and $F^{}_g=1\to F^{}_e=2$ ($\delta$ is the detuning from the transition $F^{}_g=1\to F^{}_e=2$): \\
(a) lineshape for $nk^{-3}_0=0.5$ and $\gamma^{}_{eg}/\gamma^{}_0=1/2$;\\
(b) lineshape for $nk^{-3}_0=3$ and $\gamma^{}_{eg}/\gamma^{}_0=1/2$, where there are two frequency intervals $\Delta_{\rm break}$, within which there is no solution of Eq.~(\ref{alpha_sum});\\
(c) lineshape for $nk^{-3}_0=3$ and broadening coefficient $A=\pi$ [see Eq.~(\ref{dip_dip})], where the AMCMBE-catastrophe disappears. \\
All calculations are done for $\Omega^{}_{D}/\gamma^{}_{0}=60$.} \label{87Rb_D1}
\end{figure}

Note that a simple two-level system in Fig.~(\ref{two-level}) corresponds to the transition $J^{}_g=0\to J^{}_e=1$, where $J^{}_g$ and $J^{}_e$ are angular momenta in the ground and excited states, respectively. In the general case of an arbitrary closed transition $J^{}_g\to J^{}_e$, the following equation should be used:
\begin{align}\label{alpha_gen}
&\alpha^2+2i\alpha=\\
&-i\frac{2J^{}_e+1}{2J^{}_g+1}\pi nk^{-3}_0 \gamma^{}_0\left\langle\frac{f^{}_M(v)}{\gamma^{}_{eg}-i\delta+(i+\alpha)\Omega^{}_D v/\bar{v}}\right\rangle_{v},\nonumber
\end{align}
instead of Eq.~(\ref{alpha_2}).

In the case of atoms with hyperfine structure, there are several closely lying transitions with resonance frequencies $\omega^{(q)}_0$ ($q=1,2,...$), which result in a more general equation:
\begin{align}\label{alpha_sum}
&\alpha^2+2i\alpha=\\
&-i\pi nk^{-3}_0 \gamma^{}_0\sum_{q}\left\langle\frac{C^{}_q\,f^{}_M(v)}{\gamma^{(q)}_{eg}-i\delta^{}_q+(i+\alpha)\Omega^{}_D v/\bar{v}}\right\rangle_{v},\nonumber
\end{align}
where $\delta^{}_q=\omega-\omega^{(q)}_0$. The weight coefficients $C^{}_q$ are determined by the quantum numbers of the corresponding hyperfine levels (total angular momentum $F$, electronic angular momentum $J$, nuclear spin $S^{}_n$) using the quantum theory of angular momentum. As an example of solving the equation (\ref{alpha_sum}), Fig.~\ref{87Rb_D1}(a) shows the spectroscopic dependencies of Re[$\alpha(\delta)$] for the D1 line in a gas of $^{87}$Rb atoms ($\lambda=795$~nm; $\gamma^{}_0/2\pi =5.75$~MHz) when scanning the laser frequency $\omega$ in the region of hyperfine transitions $F^{}_g=1\to F^{}_e=1$ and $F^{}_g=1\to F^{}_e=2$ (where $F^{}_g$ and $F^{}_e$ are the total angular momenta of hyperfine levels in the ground and excited states, respectively). In this case, we have the weight factors $C^{}_1=1/16$ for the transition $F^{}_g=1\to F^{}_e=1$ and $C^{}_2=5/16$ for the transition $F^{}_g=1\to F^{}_e=2$.

\section{The atomic-motion-conditioned catastrophe for the Maxwell-Bloch equations}

\begin{figure}[t]
\centerline{\scalebox{1.0}{\includegraphics{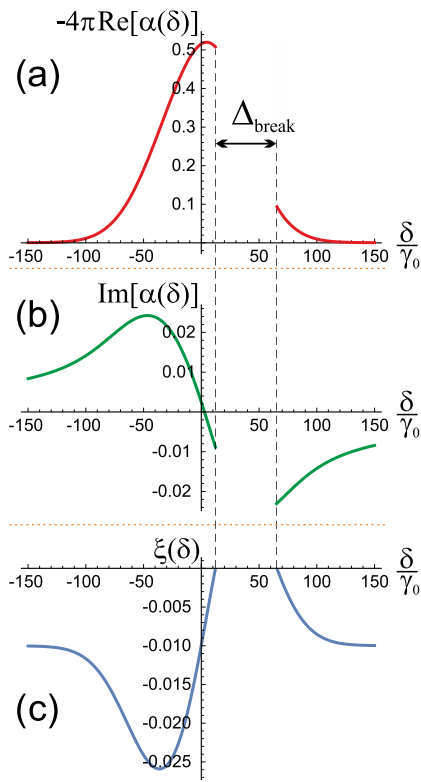}}}\caption{Spectroscopic dependence $\alpha(\delta)$ in the case where there is a frequency interval $\Delta_{\rm break}$, within which there is no solution of Eq.~(\ref{alpha_2}): (a) Doppler absorption lineshape Re[$\alpha(\delta)$]; (b) correction to the dispersion law Im[$\alpha(\delta)$]; (c) frequency dependence $\xi(\delta)$ [see Eq.~(\ref{xi})]. \\
The calculations correspond to the following parameters: $nk^{-3}_0=0.25$; $\gamma^{}_{eg}/\gamma^{}_0=1/2$; $\Omega^{}_{D}/\gamma^{}_{0}=50$.} \label{Break}
\end{figure}

In addition to the above, we have found an extremely unexpected result, where there exists no solution to Eq.~(\ref{alpha_2}) within a frequency interval $\Delta_{\rm break}$ starting from some value $nk^{-3}_0$ [i.e., only the trivial solution $E=0$ takes place for the Maxwell-Bloch equations (\ref{Max_eq})-(\ref{P})]. This is clearly seen in Figs.~\ref{Break}(a) and (b), as a discontinuity in the functional dependence $\alpha(\delta)$. Numerical analysis shows that when $\gamma^{}_{eg}/\Omega^{}_D\ll 1$ such a frequency interval exists (i.e., $\Delta_{\rm break}\neq 0$) under the condition
\begin{equation}\label{no_exist}
nk^{-3}_0>0.27\,\gamma^{}_{eg}/\gamma^{}_0\,.
\end{equation}
In the case of purely spontaneous relaxation ($\gamma^{}_{eg}=\gamma^{}_0/2$), this corresponds to $nk^{-3}_0>0.135$. Note also that the absence of a solution of Eq.~(\ref{alpha_2}) is due only to the atomic motion, because the solution (\ref{alpha_D2}) for the simplified equation (\ref{alpha_D}) exists for any set of values \{$nk^{-3}_0$, $\gamma^{}_0$, $\gamma^{}_{eg}$, $\delta$, $\Omega^{}_D$\}.

The possibility of discontinuities in $\alpha(\delta)$ is mathematically substantiated in Appendix A. In particular, under the condition
\begin{equation}\label{ratio}
\frac{3\pi nk^{-3}_0 \gamma^{}_0}{\Omega^{}_D }\ll 1\,,
\end{equation}
the existence of the solution for Eq.~(\ref{alpha_2}) additionally requires the inequality
\begin{equation}\label{xi}
\xi ={\rm Im}\left\{ \frac{\gamma^{}_{eg}-i\delta}{(i+\alpha)\Omega^{}_D}\right\}< 0\,.
\end{equation}
Moreover, at the points of discontinuity, $\xi =0$. Since Re$[\alpha (\delta)]<0$, the inequality (\ref{xi}) always holds for $\delta <0$. Therefore, in the case of the condition (\ref{ratio}), the location of the interval $\Delta_{\rm break}$ should be expected in the opposite area, i.e. for $\delta> 0$.

Figure \ref{Break} shows numerical calculations that confirm the above reasoning. As seen in Fig.~\ref{Break}(c), in the region where the solution of Eq.~(\ref{alpha_2}) exists, the condition $\xi <0$ is satisfied [see Eq.~(\ref{xi})], while at the boundaries of $\Delta_{\rm break}$ we have $\xi =0$. Moreover, we have verified that these results hold for other velocity distributions $f(v)$ (e.g., for the stepwise and Lorentzian). Note also that the absence of a solution of Eq.~(\ref{alpha_2}) cannot be explaned by the linear approximation in the electric field $E$ for the Bloch equations (\ref{Bloch_eq}), while the nonlinear contributions (in $E$) will solve this problem. Indeed, we can always choose the initial amplitude $E^{}_0$ small enough that the linear approximation in the field $E$ for the Bloch equations (\ref{Bloch_eq}) is quite correct.

\begin{figure}[t]
\centerline{\scalebox{1.}{\includegraphics{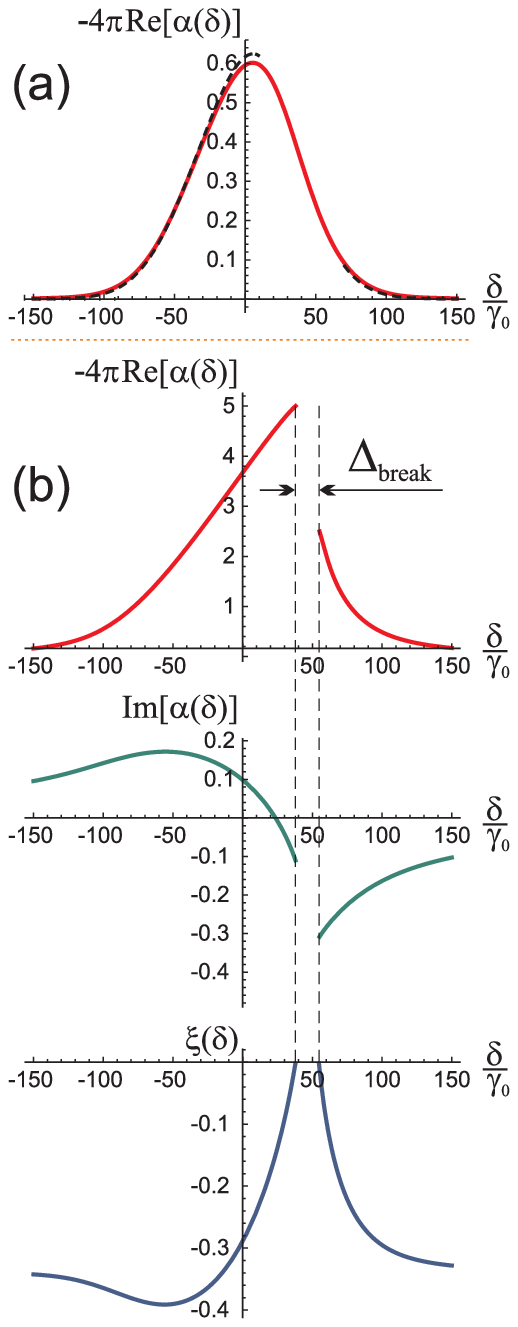}}}\caption{Spectroscopic dependence $\alpha(\delta)$ in the presence of broadening (\ref{dip_dip}) with proportionality coefficient $A=5.5$ for different atomic densities $n$:\\
(a) Doppler absorption Re[$\alpha(\delta)$] for $nk^{-3}_0=0.3$ at $A=5.5$ (solid red line) and at $A=0$ (dashed black line, where a break in the dependence is seen in the region $\delta>0$);\\
(b) frequency dependencies Re[$\alpha(\delta)$], Im[$\alpha(\delta)$], $\xi(\delta)$ in the case of $nk^{-3}_0=3$ and $A=5.5$, where there is a frequency interval $\Delta_{\rm break}$, within which there is no solution of Eq.~(\ref{alpha_2}). \\
All calculations are done for $\Omega^{}_{D}/\gamma^{}_{0}=50$.} \label{Coll_broad}
\end{figure}

As follows from Fig.~\ref{87Rb_D1}(b), in the general case of hyper-fine structure, there may be several frequency intervals $\Delta_{\rm break}$ for which there is no solution of Eq.~(\ref{alpha_sum}). Note that due to the presence of several resonant terms on the right-hand side of Eq.~(\ref{alpha_sum}), the condition for the existence of a solution will differ from Eq.~(\ref{xi}).

Logically, the absence of a solution to Eq.~(\ref{alpha_2}) can be interpreted in two ways. One, if we assume that this fact has a physical meaning, then we can expect to experimentally observe some features in the frequency interval $\delta \in \Delta_{\rm break}$, when only the trivial self-consistent solution to the Maxwell-Bloch equations exists, $E=0$. In this case, the situation looks like that the atomic medium ``does not let in'' light, which may be manifested as a total reflection of light from the boundary of the atomic cell.

Two, if the absence of the solution of Eq.~(\ref{alpha_2}) has no physical meaning, then this is a serious mathematical problem, where for a certain range of parameters the Maxwell-Bloch equations system (\ref{Max_eq})-(\ref{P}) has no solution at all. In this case, we call this problem the atomic-motion-conditioned catastrophe for Maxwell-Bloch equations (AMCMBE-catastrophe). Alternatively, the problem may be related to the mean-field approximation, which we call the atomic-motion-conditioned catastrophe of the mean-field approximation (AMCMF-catastrophe). In any case, this mathematical problem concerns the basic principles of the theoretical description of light propagation in a gas and requires further research.

In particular, if the mean free path of an atom is much larger than the wavelength $\lambda$, we can combine (at least phenomenologically) our approach with collective effects due to the interatomic dipole-dipole interaction using the following replacement in Eqs.~(\ref{alpha_2}), (\ref{alpha_gen}) and (\ref{alpha_sum}):
\begin{equation}\label{dip_dip}
\gamma^{}_{eg}=\gamma^{}_0/2+Ank^{-3}_0 \gamma^{}_0\,,\quad \delta\to\delta-Bnk^{-3}_0 \gamma^{}_0\,,
\end{equation}
where $Ank^{-3}_0 \gamma^{}_0$ and $Bnk^{-3}_0 \gamma^{}_0$ are the broadening and shift (both induced by the dipole-dipole interaction), respectively. The proportionality coefficients $A$ and $B$ are determined theoretically or experimentally. In this case, the total shift of the Doppler lineshape is the sum of the shift due to atomic-motion-induced effects [i.e., as a result of the solution of Eqs.~(\ref{alpha_2}), (\ref{alpha_gen}), (\ref{alpha_sum})] and the value $Bnk^{-3}_0 \gamma^{}_0$. Note that for a more detailed theory, we can also assume a velocity dependence for the coefficients $A(v)$ and $B(v)$ in Eq.~(\ref{dip_dip}) (e.g., $A(v),B(v)\propto f^{}_M(v)$ in the case of $\gamma^{}_{eg}/\Omega^{}_D\ll 1$).

Using above approach, we have found that the homogeneous broadening $Ank^{-3}_0 \gamma^{}_0$ in Eq.~(\ref{dip_dip}) can noticeably affect the Doppler lineshape and the regime where there is no solution to Eqs.~(\ref{alpha_2}),(\ref{alpha_gen}), (\ref{alpha_sum}) [e.g., see Fig.~\ref{Coll_broad}(a)]. For example, if we use the known values $A=\pi$ and $A=\pi\sqrt{2}$ for the D1 and D2 lines of alkali atoms (see in Ref.~\cite{Lewis_1980}), respectively, then from numerical calculations we see that the AMCMBE-catastrophe problem disappears [e.g., see Fig.~\ref{87Rb_D1}(c)] (i.e., the solution of Eq.~(\ref{alpha_sum}) exists for any value of $nk^{-3}_0$). Some difficulty remains only for a simple two-level atom, because the value $A=5.4$-5.7 (see Refs.~\cite{Kazantsev_1967,Vdovin_1967,Lewis_1980}) is not enough to completely exclude the AMCMBE-catastrophe regime [see Fig.~\ref{Coll_broad}(b)]. In the case of transition $J^{}_g=0\to J^{}_e=1$, our calculations  show total disappearance of AMCMBE-catastrophe only for $A>6.9$. A possible solution of this ``conundrum'' is to take into account the reabsorption of spontaneously emitted photons \cite{Fleischhauer_1999}, which can make an additional contribution to the broadening coefficient $A$ such that $A>6.9$. Thus, the use of Eq.~(\ref{dip_dip}) is one of possible methods to solve the AMCMBE-catastrophe problem.  However, as of now we still cannot completely exclude the presence of the physical meaning for the AMCMBE-catastrophe regime \cite{comment}.

In addition, our calculations show that taking into account the homogeneous broadening $Ank^{-3}_0 \gamma^{}_0$ in Eq.~(\ref{dip_dip}) has small effect on the magnitudes of the shifts $\bar{\delta}^{}_{\rm Re}$  and $\bar{\delta}^{}_{\rm Im}$ in Eqs.~(\ref{delta_D})-(\ref{delta_Im}) (at the percentages level for $A=5.7$) in the case of $nk^{-3}_0 <1$ and significant Doppler broadening ($\Omega^{}_D\gg \gamma^{}_{eg}$).

\section{Conclusion}

We have developed a theory of the Doppler-broadened lineshape in an atomic gas, based on the previously unknown self-consistent solution to the Maxwell-Bloch equations in the mean-field approximation and one-atomic density matrix. The effects nonlinear in the atomic density caused by the free motion of atoms were found,  which affect the lineshape (shift, asymmetry, broadening). It was shown that in the regime of significant Doppler broadening, these effects can exceed by more than one order of magnitude the contribution of interatomic dipole-dipole interactions (e.g., Lorentz-Lorenz shift). Moreover, in certain area of parameters, we have found that there exists a frequency interval $\Delta ^{}_{\rm break}$ with only the trivial self-consistent solution to the Maxwell-Bloch equations (i.e., only $E=0$). This problem (called by us AMCMBE- or AMCMF-catastrophe) concerns the basic theoretical description of light propagation in a gas and, therefore, requires further research. In particular, we have shown that this problem disappears when we take into account the homogeneous broadening due to atom-atom interaction.

Note that in numerical simulations, inhomogeneous Doppler broadening is often described using a stochastic model of motionless atoms, where the resonance frequency of each atom in an ensemble is shifted by a Gaussian-distributed random variable with zero mean and the rms value $\Omega^{}_D$. However, it should be emphasized that in this model the above-described atomic-motion-induced effects cannot be taken into account, because these effects are rigorously based on the presence of the differential operator $({\bf v}\cdot\nabla)$ in the Bloch equations for the density matrix of moving atoms.

Thus, in a resonant gas medium, the physical picture of spectroscopic effects nonlinear in atomic density is a complicated mixture of both effects due to the interatomic interaction and atomic motion. The obtained results are important for laser spectroscopy, atomic clocks and fundamental physics.

The possibilities of experimental verification of the presented results are discussed in Appendix B, where we estimate the conditions for observing the central part of the lineshape in a wide range of atomic density parameter $nk^{-3}_{0}$ for a transmission signal $I(\delta,{\cal L})$ through an atomic cell of length ${\cal L}$.

We thank I. M. Sokolov, V. L. Velichansky, and J. W. Pollock for useful discussions and comments.

\appendix
\section{}

We will consider the basic equation (\ref{alpha_2}). From a mathematical viewpoint, it is possible to assume the existence of discontinuity points in the analytical dependence $\alpha(\delta)$ from the following considerations. Let us represent the integral in the right-hand side of Eq.~(\ref{alpha_2}) as follows:
\begin{eqnarray}\label{integral}
&&\left\langle\frac{f^{}_M(v)}{\gamma^{}_{eg}-i\delta+(i+\alpha)\Omega^{}_D v/\bar{v}}\right\rangle_{v}=\nonumber\\ 
&&\frac{1}{(i+\alpha)\Omega^{}_D}\left\langle\frac{f^{}_M(v)}{\frac{\gamma^{}_{eg}-i\delta}{(i+\alpha)\Omega^{}_D}+ v/\bar{v}}\right\rangle_{v}.
\end{eqnarray}
In this case, the fraction in the integrand can be rewritten in the form
\begin{equation}\label{frac}
\frac{1}{\frac{\gamma^{}_{eg}-i\delta}{(i+\alpha)\Omega^{}_D}+ v/\bar{v}}=\frac{1}{i\xi+ \theta}=-i\frac{\xi}{\xi^2+ \theta^2}+\frac{\theta}{\xi^2+ \theta^2}\,,
\end{equation}
where
\begin{equation}\label{et}
\xi=\text{Im}\left\{ \frac{\gamma^{}_{eg}-i\delta}{(i+\alpha)\Omega^{}_D}\right\}\,,\quad \theta=v/\bar{v}+\text{Re}\left\{ \frac{\gamma^{}_{eg}-i\delta}{(i+\alpha)\Omega^{}_D}\right\}.
\end{equation}
Then, in the case of $\xi\rightarrow 0$ for Eq.~(\ref{frac}), we have
\begin{equation}\label{d}
\frac{\xi}{\xi^2+ \theta^2}\rightarrow \pi\,\text{sign}[\xi]\,\delta_{\rm Dirac}(\theta) \,,
\end{equation}
where $\delta_{\rm Dirac}(...)$ denotes the Dirac $\delta$-function. Therefore, in principle, one can expect a violation of the analyticity (for example, a discontinuity) of the function $\alpha(\delta)$ at the points where $\xi=0$ (due to the non-analyticity of the function $\text{sign}[\xi]$ at the point $\xi=0$).

\begin{figure}[t]
\centerline{\scalebox{0.5}{\includegraphics{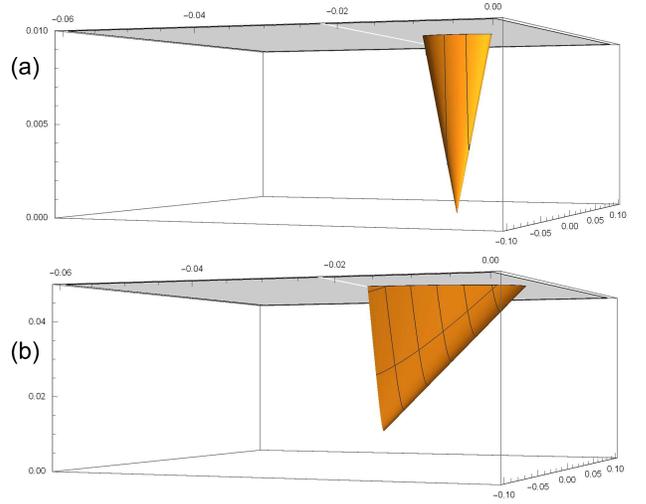}}}\caption{
The dependence of the absolute value $|F(x+iy)|$ on the complex variable $\alpha =x+iy$ in the plane ($x,y$):\\
(a) the case when a point for which $|F(x+iy)|=0$ exists, $\{nk^{-3}_0=0.1;\,\Omega^{}_D/\gamma^{}_0=50;\,\gamma^{}_{eg}/\gamma^{}_0=0.5;\,\delta/\gamma^{}_0=20 \}$;\\
(b) the case when a point for which $|F(x+iy)|=0$ does not exist, $\{nk^{-3}_0=0.2;\,\Omega^{}_D/\gamma^{}_0=50;\,\gamma^{}_{eg}/\gamma^{}_0=0.5;\,\delta/\gamma^{}_0=20 \}$.
} \label{Supl}
\end{figure}

Let us show that if the following condition
\begin{equation}\label{ratio}
\frac{3\pi nk^{-3}_0 \gamma^{}_0}{\Omega^{}_D }\ll 1
\end{equation}
is satisfied [see Eq.~(26) in the main text], the presence of a root of the equation (\ref{alpha_2}) must be accompanied by the inequality
\begin{equation}\label{xi_A}
\xi =\text{Im}\left\{ \frac{\gamma^{}_{eg}-i\delta}{(i+\alpha)\Omega^{}_D}\right\}< 0\,.
\end{equation}
Indeed, using the expressions (\ref{integral}) and (\ref{frac}), we can rewrite Eq.~(\ref{alpha_2}) in the following form:
\begin{equation}\label{alpha_3}
(\alpha^2+2i\alpha)(i+\alpha)=-\frac{i3\pi nk^{-3}_0 \gamma^{}_0}{\Omega^{}_D}\left\langle\frac{f^{}_M(v)}{i\xi +\theta}\right\rangle_{v}.
\end{equation}
Obviously, if the condition Eq.~(\ref{ratio}) holds, $|\alpha|\ll 1$ holds also. Then, discarding the small terms $\alpha^2$ and $\alpha^3$ in the right-hand side of Eq.~(\ref{alpha_3}), we obtain an approximate equation
\begin{equation}\label{alpha_4}
-2\alpha \approx -\frac{i3\pi nk^{-3}_0 \gamma^{}_0}{\Omega^{}_D}\left\langle\frac{f^{}_M(v)}{i\xi +\theta}\right\rangle_{v},
\end{equation}
whence follows
\begin{equation}\label{alpha_Re}
2{\rm Re}[\alpha] \approx \frac{3\pi nk^{-3}_0 \gamma^{}_0}{\Omega^{}_D}\left\langle\frac{\xi f^{}_M(v)}{\xi^2 +\theta^2}\right\rangle_{v}.
\end{equation}
Because in our case the wave propagates from left to right, the physical solution must satisfy the condition ${\rm Re}[\alpha]<0$, i.e., $\xi <0$. Thus, Eq.~(\ref{alpha_Re}) leads to the inequality (\ref{xi_A}), which corresponds to Eq.~(\ref{xi}) in the main text.

Note that it is possible to easily check the presence/absence of a solution for Eq.~(\ref{alpha_2}) using a graphical method. To do this, we define the complex function $F(\alpha)$ as follows:
\begin{equation}\label{Fa}
F(\alpha)=\alpha^2+2i\alpha+i\left\langle\frac{3\pi nk^{-3}_0 \gamma^{}_0f^{}_M(v)}{\gamma^{}_{eg}-i\delta+(i+\alpha)\Omega^{}_D v/\bar{v}}\right\rangle_{v}.
\end{equation}
Now you can build a three-dimensional graph of the dependence of the absolute value $|F(x+iy)|$ on the complex variable $\alpha =x+iy$ in the plane ($x,y$) for fixed values $\{nk^{-3}_0 \gamma^{}_0;\,\Omega^{}_D;\,\gamma^{}_{eg};\,\delta \}$. In this case, the existence of a solution to the equation (\ref{alpha_2}) corresponds to the presence of a point for which $|F(x+iy)|=0$. Indeed, Fig.~\ref{Supl}(a) shows the dependence $|F(x+iy)|$ in the case when such a point exists. However, if we take another set of parameters $\{nk^{-3}_0 \gamma^{}_0;\,\Omega^{}_D;\,\gamma^{}_{eg};\,\delta \}$, for which the solution of the equation (\ref{alpha_2}) is absent, then the strict inequality $|F(x+iy)|>0$ always holds, as can be seen in Fig.~\ref{Supl}(b).

\section{}

Let us consider experimental verification of the results obtained above when scanning the frequency inside the Doppler absorption line, $|\delta|<2\Omega^{}_{D}$, in the transmission signal. First of all, the initial amplitude of the field $E^{}_0$ at the entrance to the medium should be small enough to satisfy the linear approximation for the Bloch equations [see Eq.~(4) in the main text], i.e., small saturation limit. In the case of the transition $J^{}_{g}=0\rightarrow J^{}_{e}=1$, which exactly corresponds to the two-level atom, this requires the following condition:
\begin{equation}\label{low_sat}
\frac{|d^{}_{eg}E^{}_0|^2}{\gamma^2_{eg}}\ll 1\,.
\end{equation}
For an optical dipole transition ($\gamma^{}_0/2\pi\sim 1$-10~MHz), this approximately corresponds to the initial intensity $I^{}_0<0.1$-1~mW/cm$^2$, and the diameter of the light beam can be arbitrary. Considering the high level of the absorption coefficient Re[$\alpha(\delta)$] (see Figs.~2-5 and Eq.~(16) in the main text), we can estimate the thickness of the atomic cell ${\cal L}<$~10~$\mu$m (depending on the wavelength $\lambda$), which allows study of the Doppler lineshape in the transmission signal right up to $nk^{-3}_0\sim 1$.

\begin{figure}[t]
\centerline{\scalebox{0.7}{\includegraphics{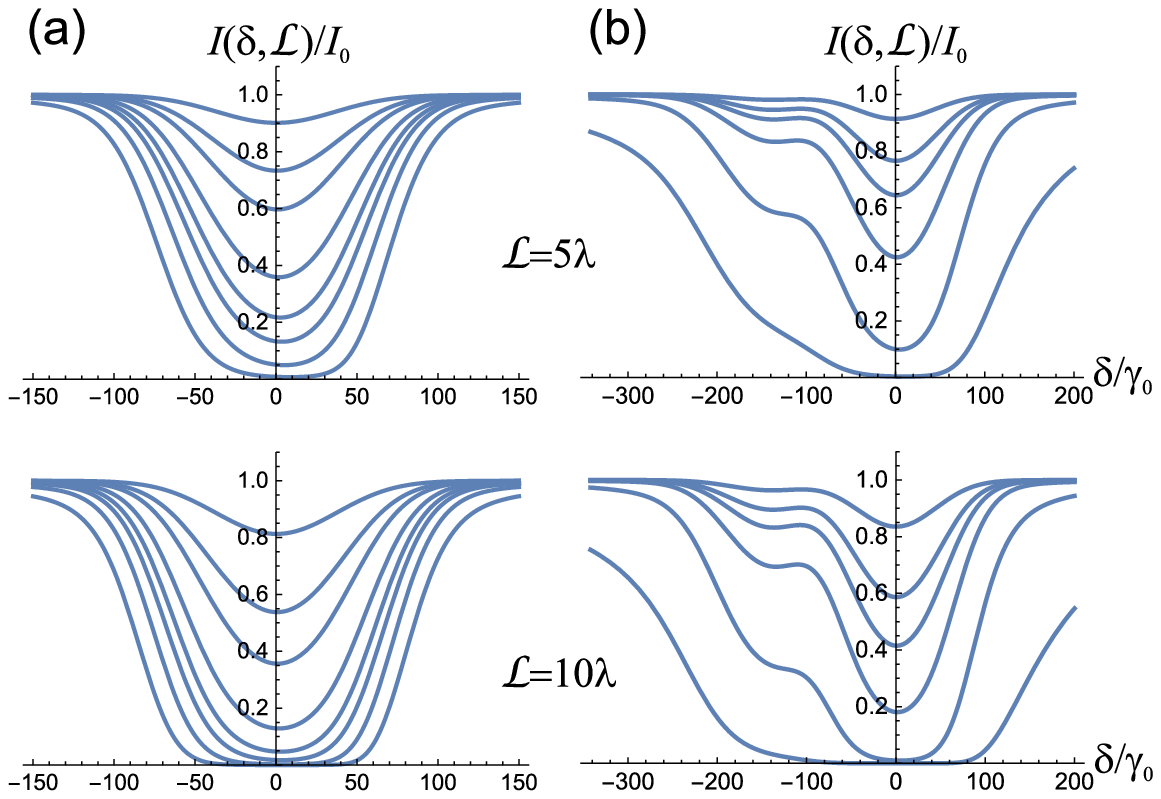}}}\caption{
Frequency dependence of the normalised transmitted intensity $I(\delta,{\cal L})/I^{}_0$ through atomic cells of length ${\cal L}$$=$$5\lambda$ and ${\cal L}$$=$$10\lambda$ for various atomic densities (from top to bottom):\\
(a) $nk^{-3}_0=0.01;\,0.03;\,0.05;\,0.1;\,0.15;\,0.2;\,0.3;\,0.5$ for two-level atom with coefficient of the homogeneous broadening $A=5.7$ (see Eq.~(28) in the main text);\\
(b) $nk^{-3}_0=0.1;\,0.3;\,0.5;\,1.0;\,3.0;\,10.0$ for the $^{87}$Rb D1 line in the region of hyperfine transitions $F^{}_g=1\to F^{}_e=1$ and $F^{}_g=1\to F^{}_e=2$ with coefficient of the homogeneous broadening $A=\pi$ (see Eq.~(28) in the main text).\\
The calculations correspond to the value $\Omega^{}_{D}/\gamma^{}_{0}=60$.
} \label{Trans}
\end{figure}

An additional difficulty lies in choosing of suitable atom with an isolated transition $J^{}_{g}=0\rightarrow J^{}_{e}=1$. The even isotopes (with zero nuclear spin) of alkaline earth atoms (e.g., Mg, Ca, Sr, Yb), which have a closed $^1$S$_0$$\to ^1$P$_1$ transition in the optical range, formally seem most convenient. However, the melting point for these elements is very high ($\sim 1000$~K), which makes it difficult to use an atomic vapor-cell. Nevertheless, for even isotopes of calcium atoms ($^{40}$Ca) vapor-cell spectroscopy is possible [A. S. Zibrov, et al., Appl. Phys. B, \textbf{59}, 327 (1994)]. Also it is possible to use even isotopes $^{196}$Hg-$^{204}$Hg of the mercury atom (melting point 234~K) with a convenient for our purposes intercombination transition $^1$S$_0$$\to ^3$P$_1$ ($\lambda=253.7$~nm, $\gamma^{}_0/2\pi = 1.3$~MHz).

As for alkali metal atoms (e.g., Rb, Cs), which are most often used in laser spectroscopy of atomic media, there always are several hyperfine degenerated states (i.e., with Zeeman sublevels). This leads to a complicated lineshape due to several closely lying Doppler absorption lines, and a redistribution of populations over the lower hyperfine states and Zeeman sublevels due to spontaneous transitions. As a result, the general condition for the linear regime of atom-light interaction has a form different from (\ref{low_sat}):
\begin{eqnarray}\label{low_sat2}
&&\gamma^{}_0\bar{\tau}\frac{|d^{}_{eg}E^{}_0|^2}{\gamma^2_{eg}}\ll 1\quad {\rm if}\quad (\gamma^{}_0\bar{\tau})>1\,, \\
&& \frac{|d^{}_{eg}E^{}_0|^2}{\gamma^2_{eg}}\ll 1\quad {\rm if}\quad (\gamma^{}_0\bar{\tau})\leq 1\,,\nonumber
\end{eqnarray}
where $\bar{\tau}$ is the flight average time of atoms through the light beam. If the beam diameter $d^{}_b$ is much less than the cell thickness ${\cal L}$, then $\bar{\tau}\sim d^{}_b/\bar{v}$. And vice versa, if $d^{}_b>{\cal L}$, then $\bar{\tau}\sim {\cal L}/\bar{v}$. Thus, for D1 and D2 lines of alkali metal atoms, the condition (\ref{low_sat2}) is satisfied for the initial intensity $I^{}_0<0.1$-1~mW/cm$^2$, if the beam diameter $d^{}_b\sim$~1-10~mm and cell thickness ${\cal L}<$~10~$\mu$m [i.e., ${\cal L}\ll d^{}_b$ and $(\gamma^{}_0\bar{\tau})\leq 1$]. Similar estimates are also valid for molecular gases with hyperfine and vibrational-rotational structures of energy levels.

To illustrate our estimates, Fig.~\ref{Trans} shows the dependences of the transmitted intensity described by the Eq.~(\ref{I}):
\begin{equation}\label{Int}
I({\delta,\cal L})=I^{}_0 e^{4\pi {\rm Re}[\alpha (\delta)]{\cal L}/\lambda}\,,
\end{equation}
for the transition $J^{}_g=0\to J^{}_e=1$ [see Fig.~\ref{Trans}(a)] and the D1 line of $^{87}$Rb [see Fig.~\ref{Trans}(b)] for various values of atomic density at atomic cell thicknesses of ${\cal L}=5\lambda$ and ${\cal L}=10\lambda$.

In the case of large detunings $|\delta|\gg \Omega^{}_D$, the linear regime of atom-light interaction is essentially simplified in relation to (\ref{low_sat})-(\ref{low_sat2}), that allows the use of thick atomic cells (${\cal L}\sim$~1-10~mm) and significantly higher light intensities. However, in our opinion, the information on possible quasi-collective effects will be noticeably leveled out in this case. In particular, the frequency interval $\Delta_{\rm break}$, within which the solution of Eq.~(\ref{alpha_2}) [or Eqs.~(23),(24)] may be absent, will be outside of the investigated frequency range.


\begin{thebibliography}{50}
\bibitem{comment}
The situation seems as follows: In order to solve the formal mathematical problem of the AMCMBE-catastrophe for the one-atom approximation and Doppler broadening ($\gamma^{}_0\ll \Omega^{}_D$), we should unexpectedly use a multi-atomic consideration that takes into account the interatomic dipole-dipole interaction. From a logical viewpoint, this looks rather strange. Indeed, in the opposite case of cold atoms, no mathematical catastrophe arises at all, because in the case of $\gamma^{}_0\gg \Omega^{}_D$ the solution of Eq.~(\ref{alpha_2}) formally exists for any value of $nk^{-3}_0$ and, therefore, there is no formal (i.e., purely mathematical) reason to take into account the atom-atom interaction. Therefore, we believe that this problem requires further careful analysis and as of now we still cannot completely exclude the existence of a physical meaning for the AMCMBE-catastrophe regime.
\end{thebibliography}
\end{document}